\documentclass{MolSpaLab}
\usepackage{graphicx}

\begin{document}
\TitreGlobal{Molecules in Space \& Laboratory}
\title{Developing A Chemical Evolutionary Sequence for Low-mass Starless Cores}
\author{\FirstName Yancy L. \LastName Shirley}
\address{Bart J. Bok Fellow, University of Arizona, 933 N. Cherry Ave, Tucson, AZ 85721}
%
%
\runningtitle{Starless Core Chemical Evolution}
\setcounter{page}{1}

\maketitle
\begin{abstract}
I review the basic processes that may be
used to develop a chemical evolutionary sequence for low-mass starless
cores.  I highlight observational
results from the Arizona Radio Observatory-Green Bank Survey.  
Observations were performed 
with the SMT 10-m, ARO 12-m, and GBT 100-m toward a sample
of 25 nearby ($D < 400$ pc) low-mass starless cores which have radiative
transfer models of the $850$ $\mu$m emission and observed SED 
($160 - 1300$ $\mu$m).  The cores were observed in the lines of
NH$_3$ (1,1) and (2,2), o-NH$_2$D $1_{11} - 1_{01}$, C$_2$S $1_2 - 2_1$,
C$_3$S $4 - 3$, HCN $1 - 0$, HC$_5$N $9 - 8$, HC$_7$N $21 - 20$,
C$^{18}$O and C$^{17}$O $2 - 1$, and p-H$_2$CO $1_{01} - 0_{00}$.  
\end{abstract}
%
\section{Introduction}
Low-mass starless cores are the earliest observed phase of isolated
low-mass star formation.  They are identified via submm
dust continuum and dense gas molecular lines, they typically contain
a few solar masses, they have sizes of approximately $0.1$ pc,
and they may form one or a few low-mass 
(M $\sim 1$ M$_{\odot}$) stars.
Several hundred starless cores have been
observed in the nearest star-forming molecular clouds and isolated Bok
globules.  
Recent large scale
surveys of nearby molecular clouds have established a remarkable connection
between the Core Mass Function and the Initial Mass Function of stars
(e.g., Motte et al. 1998, Johnstone et al. 2000), indicating the importance
of constraining the evolution of starless cores in order
to understand the initial conditions of disk and protostar formation.
Theoretically, the basic core formation and evolution process is still 
debated between a turbulent-dominated (Mac Low \& Klessen 2004) 
or ambi-polar diffusion-dominated model (Shu et al. 1987).  
Observationally, a fundamental challenge is to determine
the evolutionary state of a starless core.  Given a set of observations,
can we determine how close a starless core is to forming a protostar?

One step toward understanding the evolutionary state of a starless
core is to determine its physical structure ($n(r)$ cm$^{-3}$ and
$T(r)$ K).  
Radiative transfer modeling of submillimeter dust continuum emission
have successfully fit the density structure with hydrostatic
configurations (Bonnor-Ebert Spheres, BESs), while the calculated 
temperature structure decreases toward
the center of the core due to attenuation of the ISRF (Evans et al. 2001).  
Since BESs may be parameterized in terms of their central density,
$n_c$ cm$^{-3}$, it is natural to think that this may be the
main evolutionary variable for starless cores.  However, detailed
molecular line observations have revealed that the chemical
structure ($X(r)$) can strongly vary among cores with similar central
densities.  Figure 1 shows the example of L1498 and L1521E, two
starless cores in Taurus which have comparable $n_c$ (Tafalla et al. 2006), 
but very different chemical structures.  L1521E is centrally peaked in C$_3$S
with weak NH$_3$ emission while L1498 is centrally peaked in NH$_3$
and heavily depleted in C$_3$S.  These observations can be explained
if the cores are evolving at different rates.  \textbf{In order to
determine the evolutionary state of a starless core, it is
necessary to map the chemical structure of the core}.

\begin{figure}[ht]
\begin{center}
\includegraphics[width=9 cm]{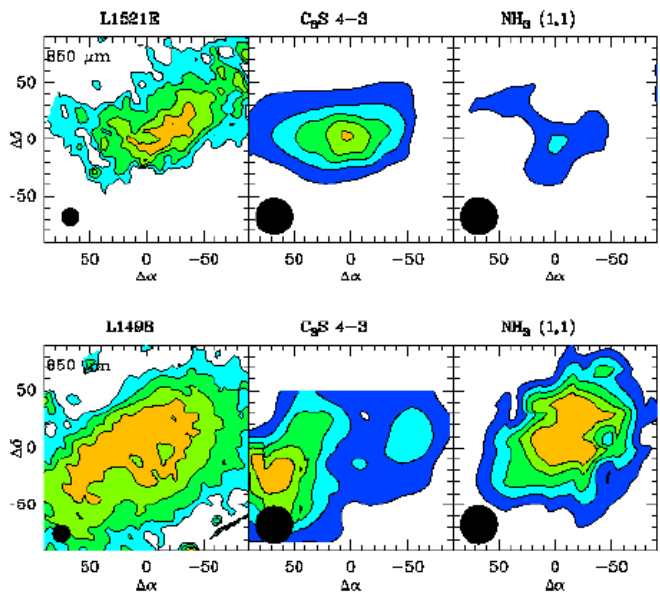}
\caption{Observations of L1521E (top) and L1498 (bottom) in 850 $\mu$m, 
C$_3$S,
and NH$_3$ showing chemical differentiation.  Data are from the ARO-GBT survey
and Shirley et al. (2005)} \label{fig}
\end{center}
\end{figure}

\section{Chemical Processes in Starless Cores}
\label{section1}

The rate at which molecules are created and destroyed differ for each
species;  therefore, the relative abundance of molecular species may be
used as a chemical clock.  A prediction that appears to be ubiquitous among
starless core chemodynamical models is that there exists
a class of molecules, named ``early-time'' molecules, which
form rapidly in the cold, moderate density environments typical of nascent
starless cores (e.g., CO, C$_2$S, C$_3$S, SO).  The early-time molecule
abundance peaks typically within a few $10^5$ years and then decreases
due to various destruction mechanisms (see below).  Another class of molecules,
named ``late-time'' molecules, build up in abundance slower and
remain in the gas phase longer at low temperatures and high densities
(e.g., N$_2$H$^+$, NH$_3$, H$_2$D$^+$).
It has been proposed that observations of the abundance ratio
of species such as [C$_2$S]/[NH$_3$] may date the chemical maturity of
a core (Suzuki et al. 1992).  Figure 1 shows an example of two cores
with very different chemical states in early-time and late-time
molecules despite having comparable central densities.

The environments of starless cores are cold ($T < 15$ K) 
and dense ($n_c > 10^4$ cm$^{-3}$), thus many gas phase species
adsorb onto dust grains.  The best example is the second most abundant
molecular species, CO, which freezes out of the gas at a rate of
$(dn_{\rm{CO}}/dt) \propto n_g T^{1/2} S n_{\rm{CO}}$,
where $n_g$ is the dust grain density and
$S$ is the sticking coefficient (Rawlings et al. 1992).  
Since the timescale for freezeout depends
on the density and temperature of the core, the amount of CO
depletion encodes the history of the physical structure of the
core. For instance, a core that evolves slowly (quasi-statically)
will have more CO depletion compared to a core that evolves quickly 
to the same $n_c$.  Complicating factors to the simple adsorption
model include competing desorption processes due to direct cosmic ray heating,
cosmic ray photodesorption, and H$_2$ formation on grains, all of which 
may be important in starless cores (Roberts et al. 2007).  CO is a destruction agent of many
gas phase ions (e.g. N$_2$H$^+$ and H$_2$D$^+$), therefore the abundance history
of these species are directly related to the amount of CO freezout
(see Figure 2).  The resulting chemical networks in heavily depleted environments
are very different (see Vastel et al. 2006).  Mapping of starless cores have revealed
a plethora of depleted species toward the dust column density peaks 
(Tafalla et al. 2006).  

Deuterium fractionation is an important chemical diagnostic in low-mass
cores.  At low temperatures, many chemical reactions involving HD favor
the formation of deuterated molecules due to the lower zero-point vibrational
energy of deuterated species compared to the hydrogenated species.  The 
classic deuteration reaction operating in the
environments of starless cores is H$_3$$^+$ + HD $\rightarrow$
H$_2$D$^+$ + H$_2$ + $230$ K, where the backreaction is inefficent
at low temperatures (Vastel et al. 2006).  As the density increases, the temperature
decreases, and species such as CO freeze-out, the deuteration of
hydrogenated species may increases up to four orders of magnitude
over the cosmic [D]/[H] $\sim 10^{-5}$ (Roberts et al. 2004).  Figure 2b illustrates 
the observed degrees of deuteration of N$_2$H$^+$
increases with the amount of CO depletion in the core (Crapsi et al. 2005).  
Similarly, since a deuterated species, such as NH$_2$D, 
may be viewed as an extreme late-time molecule, a comparison between deuterated  
and early-time molecules from the ARO-GBT survey indicate increasing
late-time vs. early-time molecules with increasing central density is
the core (Figure 2a).  

The chemical structure of the core is also important for determining
which species are good probes of the kinematical structure of
the core.  Several species (CS, H$_2$CO, HCO$^+$, HCN)
have been identified as infall tracers, molecules that are moderately 
optically thick and display asymmetric, self-aborbed line profiles.  
Recent surveys have attempted to search for evidence of large 
scale infall (e.g. Sohn et al. 2007). 
Furthermore, the linewidths of 
heavy species that lack hyperfine structure, such as C$_3$S (68 amu), 
are dominated by non-thermal motions and can trace turbulent motions
or large-scale kinematical motions along different lines-of-sight in the core.

A more thorough review of the chemical processes in starless cores
may be found in Di Francesco et al. (2005 and references therein).  
The comparison of early-time vs. late-time
species, the amount of depletion, the amount of deuteration, and 
the identification of large scale kinematical motions should be used together to 
elucidate the evolutionary state of a starless core.  Other indicators,
such as the ortho-to-para evolution of symmetric molecules with non-zero spins
and variations in
the structure of ionization fraction should also be explored.
\begin{figure}[ht]
\begin{center}
\includegraphics[width=9 cm]{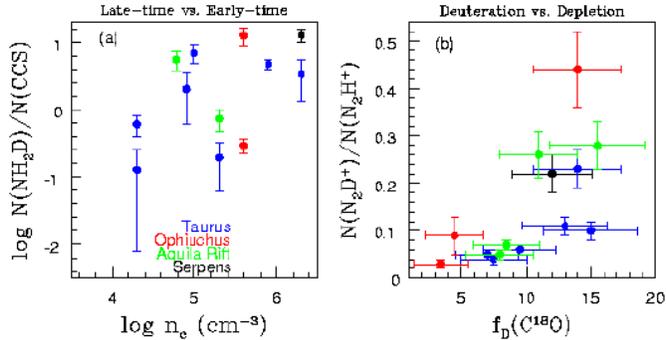}
\caption{Left: The ratio of the ``late-time'' species NH$_2$D to the ``early-time'' 
species C$_2$S vs. $n_c$ toward cores where both lines were detected in the 
ARO-GBT survey. Right: The deuterium fraction of N$_2$H$^+$ vs. the CO depletion factor
reported by Crapsi et al. (2005).} \label{fig}
\end{center}
\end{figure}

\section{Developing an Evolutionary Sequence}
\label{section2} 

Ultimately, to determine the evolutionary state of a starless core, 
we should model the molecular line radiaitve transfer
of each transition along multiple lines-of-sight
and compare to a grid of chemodynamical models (e.g. Lee et al. 2004).
This processes is computationally intensive and the current generation
of chemodynamical models have not fully explored the parameter space 
of possible conditions in nearby starless cores. 
An alternative first step is to develop a Boolean evolultionary comparison.  
A flag of $1$ (more evolved) or $0$ (less evolved) is given to a particular 
observed property of the core if the chemical criterion is met, and the sum
of flags represents the observed chemical maturity of the core.
This strategy was implemented by Crapsi et al. (2005) for a sample of
$12$ starless cores and is being extended to the sample of $25$ cores
with a larger set of chemical criteria in the ARO-GBT survey (Shirley et al.
2008, in prep.).   

While there is still much work to develop a
detailed understanding of the chemical processes within low-mass starless cores, 
it is now possible and necessary to synthesize the chemical evolutionary indicators.



\end{document}